\documentclass[preprint]{aastex63}
\usepackage{lineno}
\usepackage{multimedia}
\usepackage{booktabs}
\usepackage{graphicx}
\usepackage{pict2e}
\usepackage{amssymb}
\usepackage{amsmath}
\usepackage{tikz}

\received{??, 2020}
\revised{??, 2020}
\accepted{??, 2020}
\submitjournal{SolPhys}
\shorttitle{Brightening Detection} 
\shortauthors{Humphries, Morgan, Kuridze}
\setwatermarkfontsize{240pt}
\watermark{DRAFT}

\newcommand{\nfrag}{$N_{frag}$}
\newcommand{\threshlo}{$T_{low}$}
\newcommand{\threshhi}{$T_{high}$}
\newcommand{\dn}{DN}

\begin{document}

\title{Detecting and characterising small-scale brightenings in solar imaging data} %and AIA
\correspondingauthor{Humphries,  Morgan}
\email{llh18@aber.ac.uk, hum2@aber.ac.uk, dak21@aber.ac.uk}

\author[0000-0002-0786-7307]{Ll\^yr Dafydd Humphries}
\affiliation{Aberystwyth University \\
Faculty of Business and Physical Sciences\\
Aberystwyth, Ceredigion, SY23 3FL, Wales, UK}

\author[0000-0002-0786-7307]{Huw Morgan}
\affiliation{Aberystwyth University \\
Faculty of Business and Physical Sciences\\
Aberystwyth, Ceredigion, SY23 3FL, Wales, UK}

\author[0000-0002-0786-7307]{David Kuridze}
\affiliation{Aberystwyth University \\
Faculty of Business and Physical Sciences\\
Aberystwyth, Ceredigion, SY23 3FL, Wales, UK}

\begin{abstract}

Observations of small-scale brightenings in the low solar atmosphere can provide valuable constraints on possible heating and heat transport mechanisms. We present a method for the detection and analysis of brightenings, and demonstrate its application to time-series imagery of the Interface Region Imaging Spectrograph (IRIS) in the extreme ultraviolet (EUV). The method is based on spatio-temporal band-pass filtering, adaptive thresholding and centroid tracking, and records an event's spatial position, duration, total brightness and maximum brightness. Spatial area, brightness, and position are also recorded as functions of time throughout the event's lifetime. Detected brightenings can fragment, or merge, over time - thus the number of distinct regions constituting a brightening event is recorded over time, and the maximum number of regions are recorded as \nfrag, which is a simple measure of an event's coherence or spatial complexity. 
A test is made on a synthetic datacube composed of a static background based on IRIS data, Poisson noise and $\approx10^4$ randomly-distributed, moving, small-scale Gaussian brightenings. Maximum brightness, total brightness, area, and duration follow power-law distributions, and the results show the range over which the method can successfully extract information. The test shows that the recorded maximum brightness of an event is a reliable measure for the brightest and most accurately detected events, with an error of 6\%. Event area, duration, and speed are generally underestimated by around 15\% and have an uncertainty of $20-30$\%. The total brightness is underestimated by 30\%, and has an uncertainty of 30\%.
Applying this detection method to real IRIS quiet-sun data spanning 19 minutes over a $54.40\arcsec\times55.23\arcsec$ field of view (FOV) yields 2997 detections. 1340 of these detections either remain un-fragmented or fragment to two distinct regions at least once during their lifetime (\nfrag$\le2$), equating to an event density of $3.96\times10^{-4}$ arcsec $^{-2}$ s$^{-1}$. The method will be used for a future large-scale statistical analysis of several quiet-sun (QS) data sets from IRIS, other EUV imagers, and other types of data including H-$\alpha$ and visible photospheric imagery.
\end{abstract}


%----------------------------------------------- Introduction section ------------------------------------------------

\section{Introduction}\label{sec:intro}

Despite large advances in observation, numerical models, and theory, the heating of the solar atmosphere remains an open question. The current question is perhaps not what are the mechanisms for heating, but rather which of the several proposed mechanisms dominate, and are different mechanisms active in different regions?
There are observational and theoretical reasons to support the bulk heating of plasma in the chromosphere, some proportion of which flows into the corona to maintain high temperatures, although the debate has not been settled \citep{aschwanden, klimchuk}.
The movement of kilogauss flux tubes creates stress and reconnection at small-scale current sheets - this is the nano-flare model of heating \citep{parker, hansteen}. Brightenings, or small flare-like events, can be observed in imaging and spectroscopic data. They occur at a huge range of energies and spatial scales, and large events occur less frequently than smaller flares. Fitting the energy distribution of observed small-scale brightenings to a power law provides an estimate of the energy available at unresolved scales \citep{crosby, hudson, lu} and may give constraints on heating mechanisms. However, published results give a range of power laws which are inconclusive \citep{nhalil, aschwanden_2002}. 

Previous works have implemented a wide variety of methods for automatically detecting brightenings.
\cite{Henriques_david} use binary mapping, detection thresholding and subsequently a detection filter based on characteristics such as length and duration. 
\cite{sekse} attempt to determine whether events in neighbouring frames belong to the same chain of events based on a frame-by-frame pixel density overlap.
Work has also been done with filtering long-time intensity trends from detections in order to determine the true brightness of white-light flares \citep{mravcova}, running central median methods for resolving fine-scale dynamics \citep{plowman}, and the inference of background data and other properties from small boxes/cubes surrounding each brightening \citep{Nelson}.
There are few instances of comprehensive, homogeneous multi-instrument studies of small-scale structures/events conducted over very large data sets that also encompass and distinguish large regions of the chromosphere/corona. 
Some examples include \cite{hou_2016}'s spectral analysis of over 2700 AR \ion{Si}{4} dots, \cite{nelson_2013}'s large study of 3570 H$\alpha$ Ellerman Bombs observations, and \cite{tian}'s 176 multi-instrument penumbral dots study.

The goal of this study is to present an efficient brightening detection and characterising method, tested on synthetic data, that will be used for a large future study.
Section 2 describes the filtering and data extraction method, using its application to synthetic data to aid the method description. 
Section 3 analyses the results of its application to synthetic data, and the optimisation of certain method parameters. 
The application of the method to IRIS observations of a QS region is also discussed, along with some preliminary results.
A summary and description of future application of the method is given in section 4. 
%Section 4 applies the method to an IRIS observations of a QS region, and presents some preliminary results. 
%------------------------------------------------- Method section -------------------------------------------------

\section{Method} \label{sec:method}

\begin{figure*}[t]
\centering
%\begin{picture}(-30,30)
\includegraphics[trim={1cm 0.5cm 1cm 0cm},clip,width=0.8\textwidth]{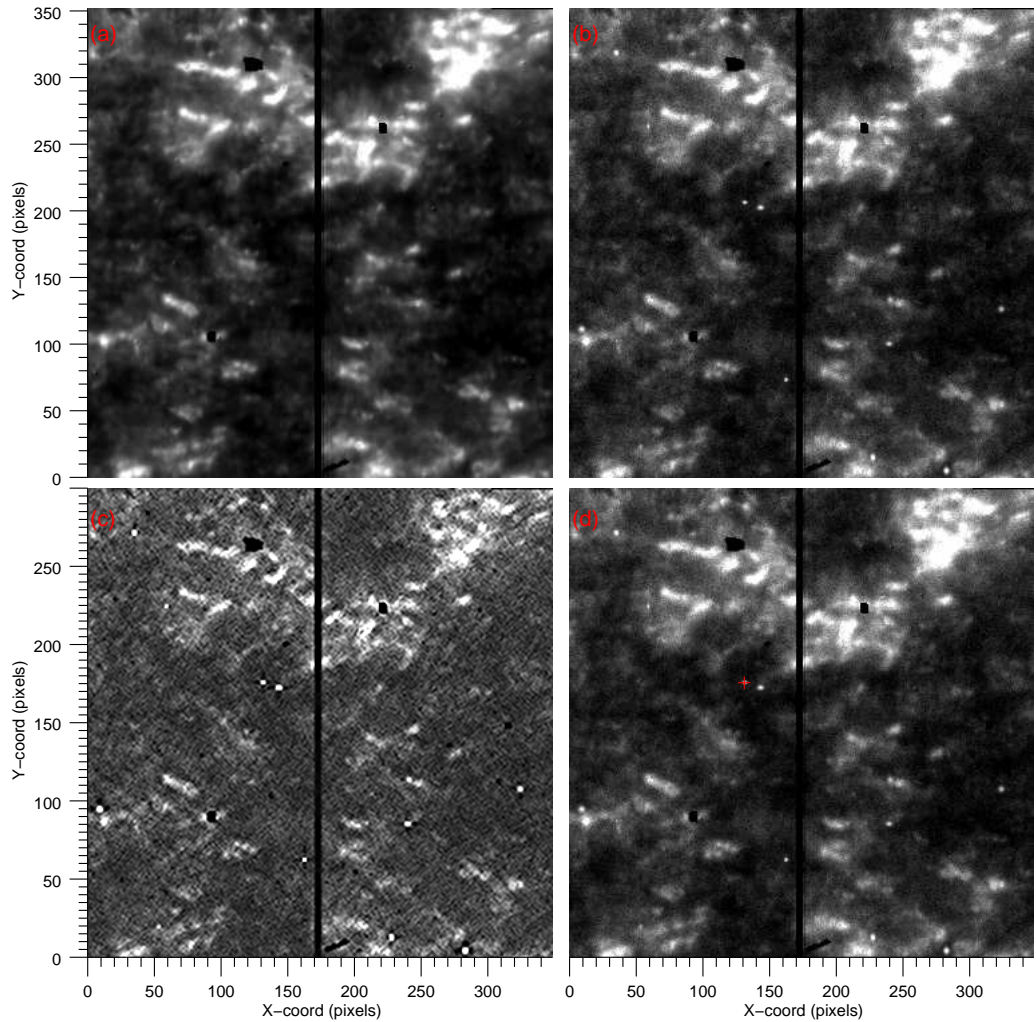}
%\put(0.8,5.1){(a)}
%\put(9.3,5.1){(b)}
%\put(4.7,0.0){(c)}
%\end{picture}
\caption{A simple representation of the creation of a synthetic datacube and the detection of a brightening therein. (a) shows a frame from the mean background of the synthetic data. 
(b) is the same as (a) but with Poisson noise and Gaussian brightenings added. 
(c) is image (b) following the filtering method. 
(d) is the same as (b) with a brightening highlighted by a red plus. The plus represents the position of maximum amplitude for this brightening.
}
\label{fig:method_comp}
\end{figure*}

\subsection{A synthetic datacube}
A datacube of synthetic data is created in order to test the method, and to facilitate the description of the method. For this purpose, this datacube is deliberately kept simple, with a low number of bright points to avoid overlap, and with properties (brightness, area, and duration) that facilitate detection. The background image for the synthetic datacube is created from a real IRIS slit-jaw observation time series. Details of the quiet-sun observations are given in section \ref{sec:results real}. We take the mean intensity over time at each spatial pixel over the whole data set in order to create a background image, shown in figure \ref{fig:method_comp}a. This image is then replicated 120 times, forming a datacube over time of size $[352, 348, 120]$, a size typical of IRIS slit-jaw datacubes, and corresponding to the real datacube presented later. For each pixel, a random time series is created from a Poisson distribution based on that pixel's mean intensity. Thus the synthetic data has a background intensity which varies randomly over time due to Poisson noise. $\sim1285$ brightenings are added to the datacube. These have Gaussian profiles of intensity in space and time, with random distributions of spatial and time position, area, and duration, some of which can be seen in figure \ref{fig:method_comp}b. The spatial $x$ and $y$ Gaussian widths range from $0.5$ to $3$ pixels (standard deviations), and their durations vary from  $1$ to $4$ time steps. These ranges are based on detected bright points in the real data.
The Gaussian amplitudes (or brightness) are also set at random, with a distribution between 10 and 30 times the standard deviation of the datacube over time. 
Random linear motions are imposed to these brightenings ranging from $0$ to $10$ pixels over the bright point's lifetime - loosely corresponding to the motions we see and detect in the real data. Figure \ref{fig:method_comp}b shows a frame from the datacube after the addition of Poisson noise and several Gaussian brightenings.
These four components - a static smoothly-changing background, Poisson noise, random Gaussian brightenings and random linear motions - provide an appropriate analogue to real QS IRIS data.\\

\subsection{Band-pass filtering}

The first step of the detection method consists of a band-pass filter in space and time.
The filtering is achieved by convolution with appropriate kernels along each dimension of the datacube. 
The kernels are formed using  IDL's \textit{DIGITAL\_FILTER} function \citep{Walraven}.
Considerable efficiency can be achieved by applying convolution to the datacube $S$ with vector kernels in the two spatial and one temporal dimension sequentially, rather than convolving once with a 3-dimensional kernel.
The filtered cube $F$ is defined as 
\begin{equation}
\begin{aligned}
S^\prime &=S \ast\alpha (x)\\
S^{\prime\prime} &=S^\prime \ast\beta (y)\\
F & = S^{\prime\prime} \ast\gamma (t) ,\\
\end{aligned}
\end{equation}
where $\alpha (x)$, $\beta (y)$ and $\gamma (t)$ are the $x$, $y$ and $t$ kernels respectively. 
Note that in this work the $x$ and $y$ kernels are identical, although there may be instances where different band-pass parameters may be appropriate (e.g. for data close to the limb). We avoid this complication in this study by choosing data that is close to disk center. 

The most important parameters for these filter kernels are the low and high frequency values controlling the band-pass, $f_{low}$ and $f_{high}$. Increasing the $f_{low}$ value reduces the power of slow temporal variation and spatially smoothly-changing features, whilst decreasing the $f_{high}$ value reduces the power of high-frequency temporal variations (i.e. background noise). Figure \ref{fig:method_comp}c shows a frame of the datacube following filtering, with $f_{low}=0.09$ and $f_{high}=0.40$. We further illustrate the filtering effect in figure \ref{fig:1d} on a slice of the datacube signal which passes through a bright point. Three slices are shown, corresponding to cuts across the $x, y, t$ dimensions. The unfiltered signal is shown as the black line. The red line shows the data following a convolution with a band-pass filter kernel with $f_{low}=0.09$ and $f_{high}=0.40$ (in units of fractions of the Nyquist frequency), showing effective dampening of the slowly changing background and high-frequency noise. The blue line shows a varying threshold parameter which is described in the following subsection: any red points above the blue threshold are detected as a candidate brightening event. The selected brightening event is made prominent by the filtering, and is clearly above the threshold. The choice of the $f_{low}$ and $f_{high}$ frequencies affects the results, and will be discussed later in the context of optimising the method's performance on the synthetic datacube.

\subsection{Thresholding for initial detection}
We estimate a threshold which is based on the Poisson noise model for each spatial pixel, thus the threshold varies over the field of view, but remains constant over time. 
Whilst the Poisson noise level is equal to the square root of the mean signal over time, we wish to define a threshold based on the filtered, rather than original, data. 
We therefore calculate a threshold numerically.
A datacube is generated in the same manner as the synthetic data above, but without the addition of bright points - so a static background with time-varying Poisson noise. 
This cube is then band-pass-filtered over all dimensions as described above.
A spatially-varying threshold (\threshhi) is then defined as the standard deviation over time, $\sigma$, of the filtered datacube multiplied by a constant. This constant is discussed in the following section.

\begin{figure*}[t]
\centering\includegraphics[trim={1cm 0.5cm 1cm 0cm},clip,width=0.8\textwidth]{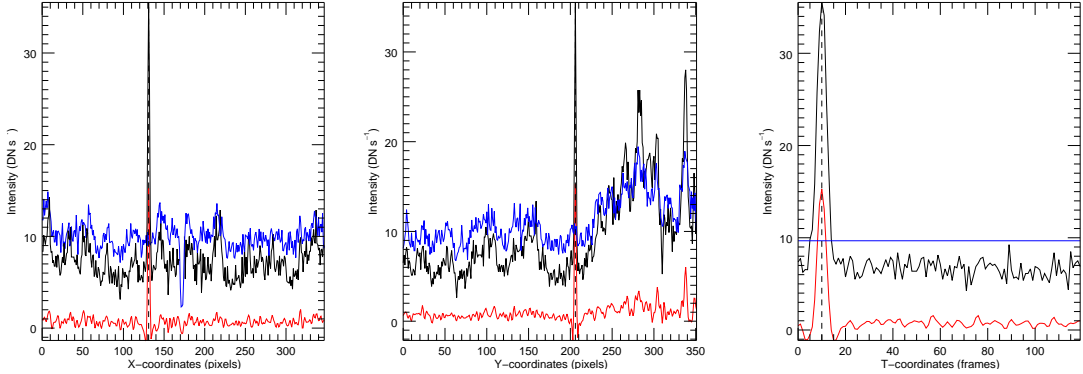}
\caption{A demonstration of the filtering and thresholding method, showing slices through the datacube passing through a bright point across the $x$ (left), $y$ (center), and time dimensions (right). Unfiltered synthetic data is shown in black. Filtered results are shown in red, and the blue line denotes a threshold value that enables detection of the narrow synthetic brightenings (see text). The horizontal dashed line shows the position of the maximum brightness of the bright point.
}
\label{fig:1d}
\end{figure*}

The red cross in figure \ref{fig:method_comp}d indicates a particular region from \ref{fig:method_comp}c that is above this threshold. This is considered as a detection, or a candidate brightening event. This event corresponds to the bright point shown in figure \ref{fig:1d}. 
Following thresholding, candidate brightenings that are smaller than 25 voxels or last less than 5 frames are discarded. For each detected event, we take the extended surrounding region and apply a lower threshold - \threshlo\ - such that any voxels within this local region that are above \threshlo\ and which contain the original bright points detected using \threshhi, are defined as the final detected bright point: thus the small region identified using \threshhi\ is grown to a larger region using \threshlo. This two-step process using an initial high threshold then a lower threshold enables effective isolation of separate bright points and increases the number of voxels which more accurately matches the total brightness of that bright point.

\begin{figure*}[t]
\begin{tikzpicture}
    \node[anchor=south west,inner sep=0] (image) at (0,0) {\includegraphics[trim={0.5cm 0.7cm 0cm 0cm},clip,width=\textwidth]{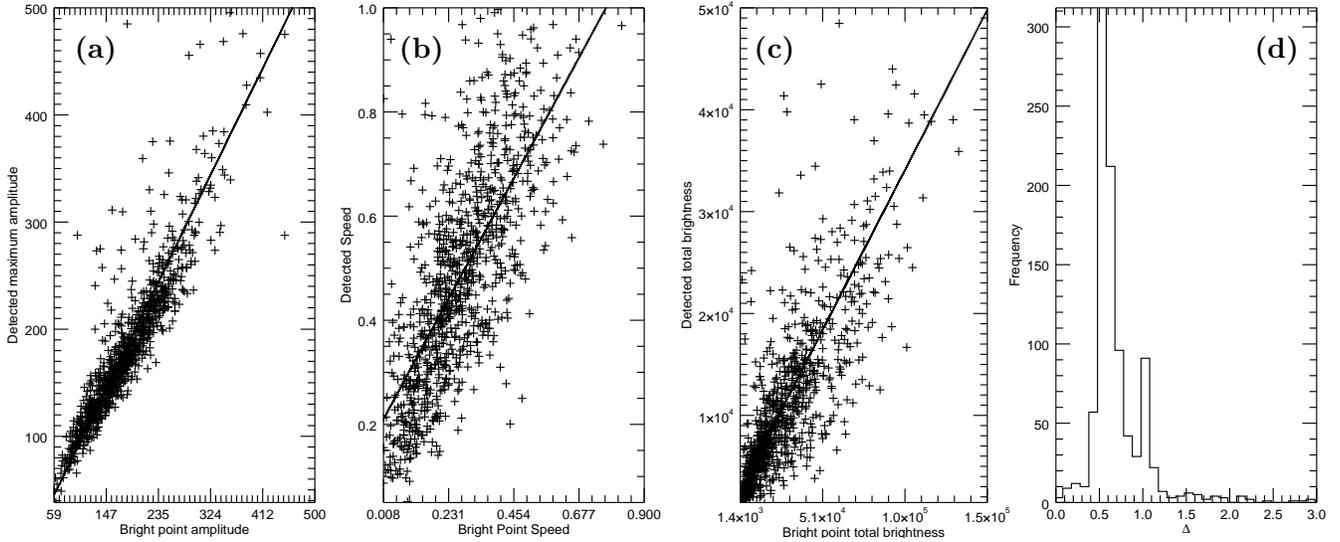}};
    \begin{scope}[
        x={(image.south east)},
        y={(image.north west)}
    ]
        \node [black, font=\bfseries] at (0.08,0.89) {(a)};
        \node [black, font=\bfseries] at (0.32,0.89) {(b)};
        \node [black, font=\bfseries] at (0.58,0.89) {(c)};
        \node [black, font=\bfseries] at (0.95,0.89) {(d)};
    \end{scope}
\end{tikzpicture}
\caption{Comparisons of detected and true properties of the synthetic bright points. (a) synthetic amplitudes vs detected brightness; (b) synthetic speed vs detected speed; (c) synthetic total brightness vs detected total brightness, and; (d) a histogram of $\Delta$ values for the event-by-event analysis.}
\label{fig:stats}
\end{figure*}

For each detected region, the following information is recorded: 
\begin{itemize}
    \item The duration of the event.
    \item The volume, defined as the number of all voxels contained in a brightening event.
    \item The background brightness, defined as the median brightness of the voxels of the extended region surrounding the bright point which are below \threshlo.
    \item The maximum brightness, or the count at the brightest voxel minus the estimated background intensity. 
    \item Total brightness, or the counts minus the background count summed across all the bright point voxels.
    \item The maximum brightness at each time step. 
    \item The total brightness at each time step.
    \item The spatial centroid of each region during each time frame, weighted by the counts at each voxel.
    \item \nfrag. Many detected brightenings in real data do not maintain a single, coherent region over time - single regions can fragment into two or more regions, or several isolated regions may merge. This is either real, a consequent of noise, is due to some parts of a brightening becoming faint and falling below the detection threshold, or is due to the complexity of the underlying `static' structure affecting the detection method. For each brightening event, the number of fragments per time step is recorded, and the maximum number of fragments over the event's lifetime is recorded as value \nfrag. For example, if an event remains unfragmented, then $N_{frag}=1$ or if an event fragments to two regions at least once during its lifetime, then $N_{frag}=2$, and so on. This is an approximate measure of how spatially coherent a brightening is over its lifetime. 
    \item Average speed. For events with \nfrag$=1$, the speed between timesteps is calculated from the spatial centroid and the observational cadence, and an average speed calculated over the event's lifetime. A more sophisticated analysis is required for \nfrag$>1$, reserved for future work.
\end{itemize}
For application to real data, \nfrag\ is useful to quickly isolate the most coherent brightenings for further study. This characteristic will be explored further in the Results \& Discussion section. We initially explored the possibility of fitting these bright point profiles to Gaussian functions in space and time. This was unsuccessful due to noise, the complexity of the bright points (particularly considering fragmentation), the high numbers of parameters needed to fully describe a moving Gaussian shape in 3 dimensions, and the small number of detected voxels for some candidate events. Furthermore, \cite{vissers} suggest that a Gaussian model may not be appropriate for all TR phenomena.

\section{Results \& Discussion} 
\subsection{Synthetic datacube} \label{sec:results synth}

\begin{deluxetable*}{cccccccccccc}
\tablecaption{Comparison of some frequency limits, threshold values and their effect on detection precision. $f_{low}=0.09$, $f_{high}=0.40$, \threshlo$=7$ and \threshhi$=9$ are chosen as the parameters for use with real IRIS data \label{tab:1}
}
\tablehead{
\colhead{$f_{low}$} & 
\colhead{$f_{high}$} & 
\colhead{\threshhi} &
\colhead{\threshlo} &
\colhead{TP} &
\colhead{FP} &
\colhead{FN} &
\colhead{TN} &
\colhead{Precision} &
\colhead{Accuracy} &
\colhead{Precise det.} & 
\colhead{Event det.}\\
\colhead{} & 
\colhead{} & 
\colhead{} & 
\colhead{} & 
\colhead{\%} &
\colhead{\%} &
\colhead{\%} & 
\colhead{\%} &
\colhead{\%} & 
\colhead{\%} & 
\colhead{$\Delta<10$} &
\colhead{\%} 
}
\startdata
0.10 & 0.50 & 5 & 5 & 29.50 & 3.32 & 70.50 & 99.96 & 89.86 & 99.19 & 100 & 79.98\\
0.09 & 0.40 & 10 & 7 & 47.30 & 8.08 & 52.70 & 99.91 & 85.40 & 99.33 & 99.59 & 76.48\\
0.09 & 0.40 & 9 & 7 & 48.51 & 8.91 & 51.49 & 99.90 & 84.48 & 99.33 & 99.23 & 80.84\\
0.10 & 0.50 & 7 & 5 & 24.84 & 2.79 & 75.16 & 99.97 & 89.91 & 99.14 & 100 & 56.54\\
0.11 & 0.40 & 5 & 4 & 44.34 & 13.92 & 55.66 & 99.85 & 76.11 & 99.23 & 100 & 87.46\\
0.11 & 0.40 & 6 & 5 & 38.29 & 7.99 & 61.71 & 99.91 & 82.73 & 99.20 & 100 & 81.85\\
0.10 & 0.50 & 5 & 3 & 45.57 & 11.33 & 54.43 & 99.87 & 80.08 & 99.28 & 100 & 79.67\\
\enddata
\tablecomments{Frequency limits are fractions of the Nyquist frequency.
}
\end{deluxetable*}

We apply the method to the synthetic datacube. The results are compared to the known distribution of the synthetic brightenings as a test of the method, and used to optimise various parameters. There are two ways to assess our method using synthetic data: a voxel-by-voxel set of statistics, and an event-by-event comparison. For the voxel-by-voxel statistics, we calculate the following:
\begin{itemize}
    \item True Positive (TP) - the number of detected voxels which correspond to synthetic brightening voxels 
    \item False Positive (FP) - the number of detected voxels which do not correspond to synthetic brightening voxels
    \item False Negative (FN) - the number of non-detected voxels that correspond to synthetic brightening voxels
    \item True Negative (TN) - the number of non-detected voxels that correspond to voxels without synthetic brightenings
\end{itemize}
Note that we define a synthetic brightening voxel as voxels that have a brightness greater than 30\%\ of their respective brightening's maximum amplitude, calculated from each brightening's Gaussian parameters. These criteria also define the Precision and Accuracy percentages, which are determined as:

\begin{equation}
Precision=\frac{TP}{TP+FP}\times100 \%, 
\end{equation}

\begin{equation}
Accuracy=\frac{TP+TN}{TP+TN+FP+FN}\times100 \%, 
\end{equation}

For an event-by-event comparison, a simple criterion we use is the distance of a detection's centroid to the closest true brightening, $\Delta=\sqrt{\Delta x^2+\Delta y^2+\Delta t^2}$. This criterion allows a percentage scoring of the method with regards to accurate event detection, defined as
\begin{equation}
\frac{n_{good}}{n_{total}}\times100 \%, 
\end{equation}
where $n_{good}$ is the number of detections with the stringent criterion $\Delta \leq 10$, and $n_{total}$ is the total number of brightenings in the synthetic datacube. Detections with $\Delta>10$ are considered spurious.

Table \ref{tab:1} shows various values of these criteria based on different values of the $f_{low}$, $f_{high}$, \threshlo\ and \threshhi\ parameters. For the synthetic data, we find an optimal band-pass at $f_{low}=0.09$, $f_{high}=0.40$ with threshold values of \threshlo$=7$ and \threshhi$=9$, whereby the detection percentage mentioned above reaches $\sim81\%$ and the TP detection reaches $\sim50\%$ while FPs remain $<10\%$ and event-by-event accuracy remains high. The rest of this work is based on these parameters.

For `true' event detections ($\Delta \leq 10$), figure \ref{fig:stats}a shows that there is a good correspondence between the detected maximum brightness and the true maximum brightness, with a mean absolute relative deviation (MARD) of $10.32$\% and a gradient of 1.13. 
Figure \ref{fig:stats}b compares the detected and true speed of the bright point, with a gradient of 1.04 and a MARD of $130.81$\%. Figure \ref{fig:stats}c shows the detected and true total brightness with a gradient of 0.32 and MARD $59.60\%$. This gradient of 0.32 shows that the total brightness is greatly underestimated. Since the maximum brightness of an event is accurately found, this underestimation is due to non-detection of brightening voxels, or an underestimation of a brightening's volume. This can be improved by decreasing \threshlo, but at the expense of increasing false positives - our choice of parameters is therefore a compromise. Figure \ref{fig:stats}d shows a histogram of $\Delta$ values, $\sim99\%$ of which are $\le3$. These comparisons give approximate estimates of the uncertainty levels which we can assign to the brightnesses of detections in real data: the uncertainty of maximum brightness amplitude is small compared to general radiometric calibration errors, whilst the uncertainty in the total brightness is large, and comparable to the calibration errors. 

All synthetic detections are dominated by $N_{frag}=1$ values, with some detections with $N_{frag}=2$. In an analysis of real data, a user can choose $N_{frag}=1$ values only if they want to exclude overlapping or dubious detections. However, as stated in section \ref{sec:method}, analysing the distribution of $N_{frag}$ gives an idea of the complexity of these events.

\begin{figure*}[t]
\begin{tikzpicture}
    \node[anchor=south west,inner sep=0] (image) at (0,0) {\includegraphics[trim={0cm 0.3cm 0cm 0.5cm},clip,width=\textwidth]{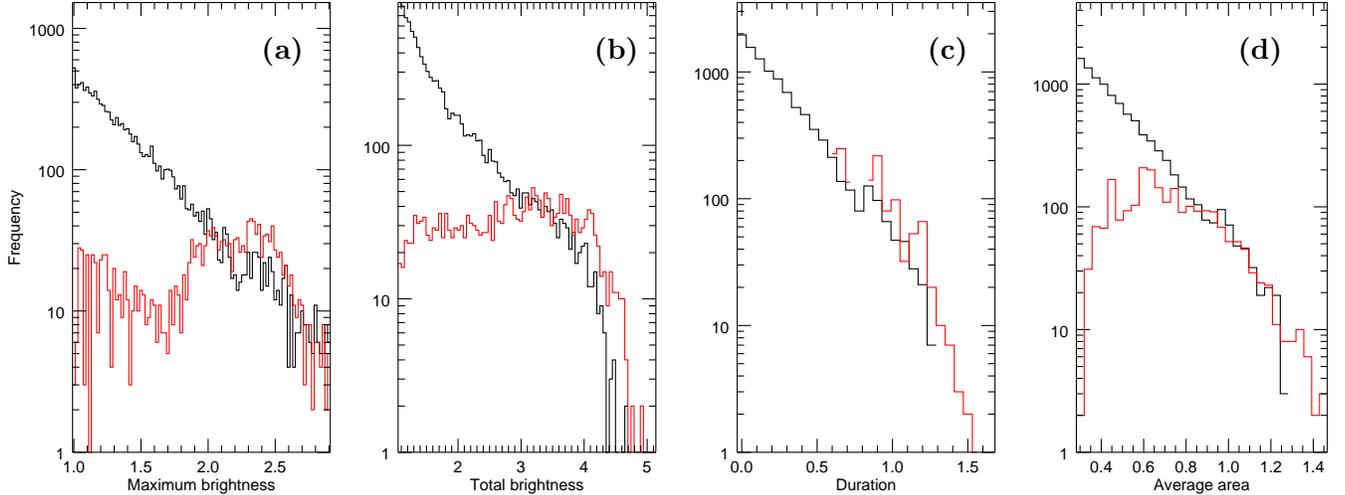}};
    \begin{scope}[
        x={(image.south east)},
        y={(image.north west)}
    ]
        \node [black, font=\bfseries] at (0.225,0.89) {(a)};
        \node [black, font=\bfseries] at (0.471,0.89) {(b)};
        \node [black, font=\bfseries] at (0.715,0.89) {(c)};
        \node [black, font=\bfseries] at (0.945,0.89) {(d)};
    \end{scope}
\end{tikzpicture}
\caption{Input (black) and detected (red) distributions of (a) Maximum brightness, (b) Total brightness, (c) Duration (measured in time steps or frames), and (d) spatial area (measured in pixels), for the 10,000-event synthetic datacube. Axes are log$_{10}$ of each characteristic.
}
\label{fig:power-law}
\end{figure*}

\subsection{A more complicated dataset} \label{section:10k}
\begin{figure*}[t]
\begin{tikzpicture}
    \node[anchor=south west,inner sep=0] (image) at (0,0) 
    {\includegraphics[width=\textwidth]{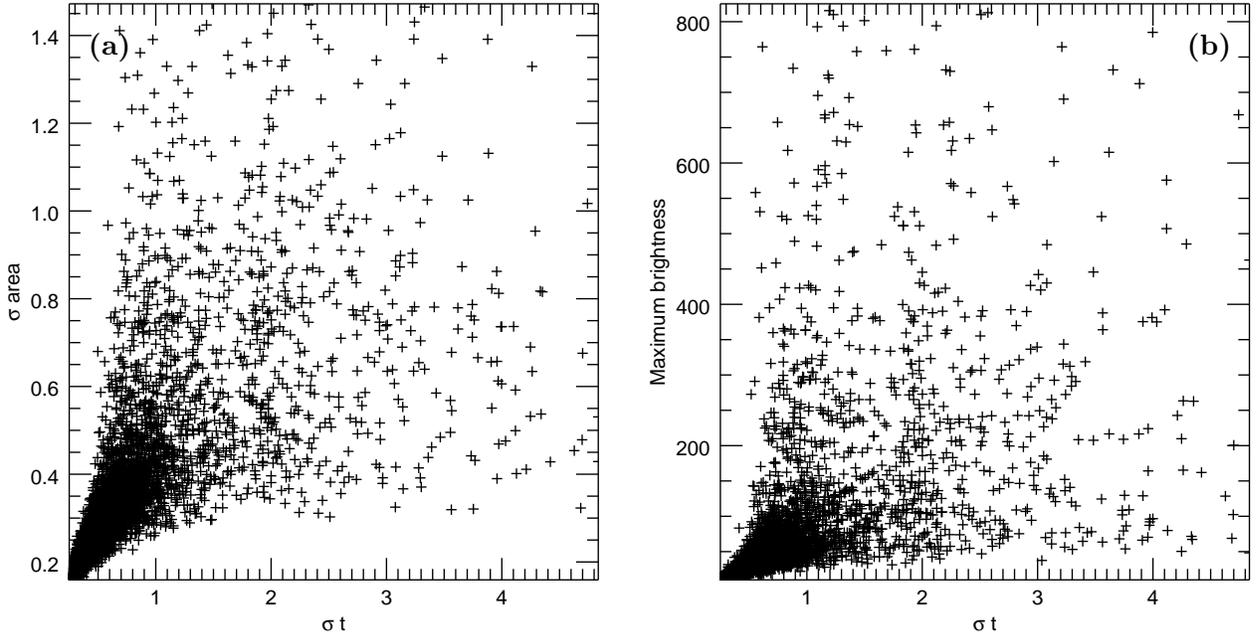}};
    \begin{scope}[
        x={(image.south east)},
        y={(image.north west)}
    ]
        \node [black, font=\bfseries] at (0.13,0.885) {(a)};
        \node [black, font=\bfseries] at (0.94,0.885) {(b)};
    \end{scope}
\end{tikzpicture}

\caption{The relationship between input values for the 10,000 brightenings, showing (a) The $1\sigma$ Gaussian area against the $1\sigma$  Gaussian time (or duration), and (b) the maximum brightness (Gaussian peak amplitude above the background) against duration.}
\label{fig:inputscatterplots}
\end{figure*}

A second synthetic datacube is created containing $\approx10^4$ brightenings. The area, duration and maximum brightness of these brightenings follow continuous power-law distributions. The lower limits of these distributions are set at values below that which we expect to detect. The lower limit for the maximum brightness amplitude of the Gaussian events is 10 \dn\ which is equal to the minimum $2\sigma$ noise level of the background. The lower limit for the spatial $1\sigma$ width is 0.4 pixels, and the lower limit for the temporal $1\sigma$ width is 0.25 time steps. The slopes of the power law distributions, as set by the standard parameter $\alpha$ (see \citep[e.g.][]{clauset}), are $\alpha_B = 2$, $\alpha_A = 3$, and $\alpha_t = 2.5$, for brightness, area and duration respectively. These $\alpha$ values are set so that the largest values of each property are at expected values. That is, so the events do not become too large or long-lived. These power laws are shown as the black lines in figure \ref{fig:power-law}. We assume that there is an approximate linear relationship between brightness, area and duration. For example, we expect the brightest events to possess, in general, a larger area and duration. We use a local randomisation process to impose these approximate relationship from the initial distributions, with the relationships shown in the plots of figure \ref{fig:inputscatterplots}. The events have a uniform distribution of speeds ranging from zero to two pixels per timesteps, in random directions, and are placed in random positions throughout the datacube. Due to their speed and position, some events may appear or disappear at the datacube edges over time, and many events may overlap. In the case of overlaps, the maximum brightening at a given pixel is retained.

This datacube is likely more representative of real-world IRIS event distributions and is a more realistic test for the detection method, whereby a large number of events are difficult or impossible to distinguish from noise, and many other events will not meet the filtering criteria i.e. some events will be too small or large, too faint, or too short-/long-lived. The same filtering process and parameters described in the previous sections are applied to this datacube. $\sim23\%$ of the 10,000 events are detected. For these detections, 88\%\ are recorded as one coherent event ($N_{frag}=1$), and 9\%\ split into two regions at least once during their detected lifetimes ($N_{frag}=2$). 

The resulting detected power laws are shown in red in figure \ref{fig:power-law}. Figures \ref{fig:power-law}a shows a reasonable agreement between the slopes of the input/detected maximum brightness distributions only above $\approx 160$\dn. Below this limit, many events are detected, but at a small fraction of the true number. This fraction decreases rapidly with decreasing brightness. Above this limit, the number of detected events is slightly larger than the true number, showing an overestimation of brightness for some events. This is likely due to overlaps between brightenings, or an underestimation of the background brightness. A similar distribution holds for the total brightness. In analysing distributions of detections in real data, this decrease in detection rate for dimmer events must be properly modelled, for example following the approach of \citet{clauset}. Without accounting for the efficiency of the method to various brightness values, the power law slope $\alpha$ will always be greatly underestimated. This will be explored further in future work.
Although detections may not directly correspond to input synthetic brightenings, the distribution of the detections still depends on the distribution of the synthetic inputs, even if the relationship becomes nonlinear.  The nature of the detection process fundamentally changes the distribution of detections in a stochastic fashion and therefore a one-to-one relationship between true events and those characterized by detection algorithms is not always possible.

%https://iopscience.iop.org/article/10.1086/308867/pdf
Additionally, this study is limited to the 1400 \AA\ channel. It is possible and indeed likely that these brightening events could be emitting in several other wavelengths, which would suggest that energy output measurement are systematically underestimated by using only one channel. Indeed, some events may only exist at higher temperatures and energies, and this will have a direct impact on power-law gradients. 
This is similar to the arguments given by \cite{aschwanden_2002} and \cite{Aschwanden_Charbonneau_2002} in the context of nanoflares at much higher temperature ranges.

The distribution of detected durations shown in figure \ref{fig:power-law}c follows the distribution of input durations well for events with $\sigma$ time width of $\approx$1 or longer. Some event's durations are overestimated - this is likely due to overlaps between brightenings. No events with $\sigma$ time widths of less than 0.6 are detected - this is due to the method discarding detections below this limit in order to reduce the number of false positives. The area distribution of figure \ref{fig:power-law}d follows the input distribution well above a limit of $\approx0.9$ pixel $\sigma$ area, with some event areas overestimated. The method does detect many small-area events below this limit, but at a fraction that decreases with decreasing area. 
While the detected distribution and input distribution match well down to a lower limit, this does not necessarily imply a one-to-one correspondence between a match and its true value. The slope of each distribution does not depend on the exact value of any detection (due to the scale-invariant nature of power-laws); only how they are distributed.

\begin{figure*}[t]
\begin{tikzpicture}
    \node[anchor=south west,inner sep=0] (image) at (0,0) {\includegraphics[trim={0 0 0 0},clip,width=\textwidth]{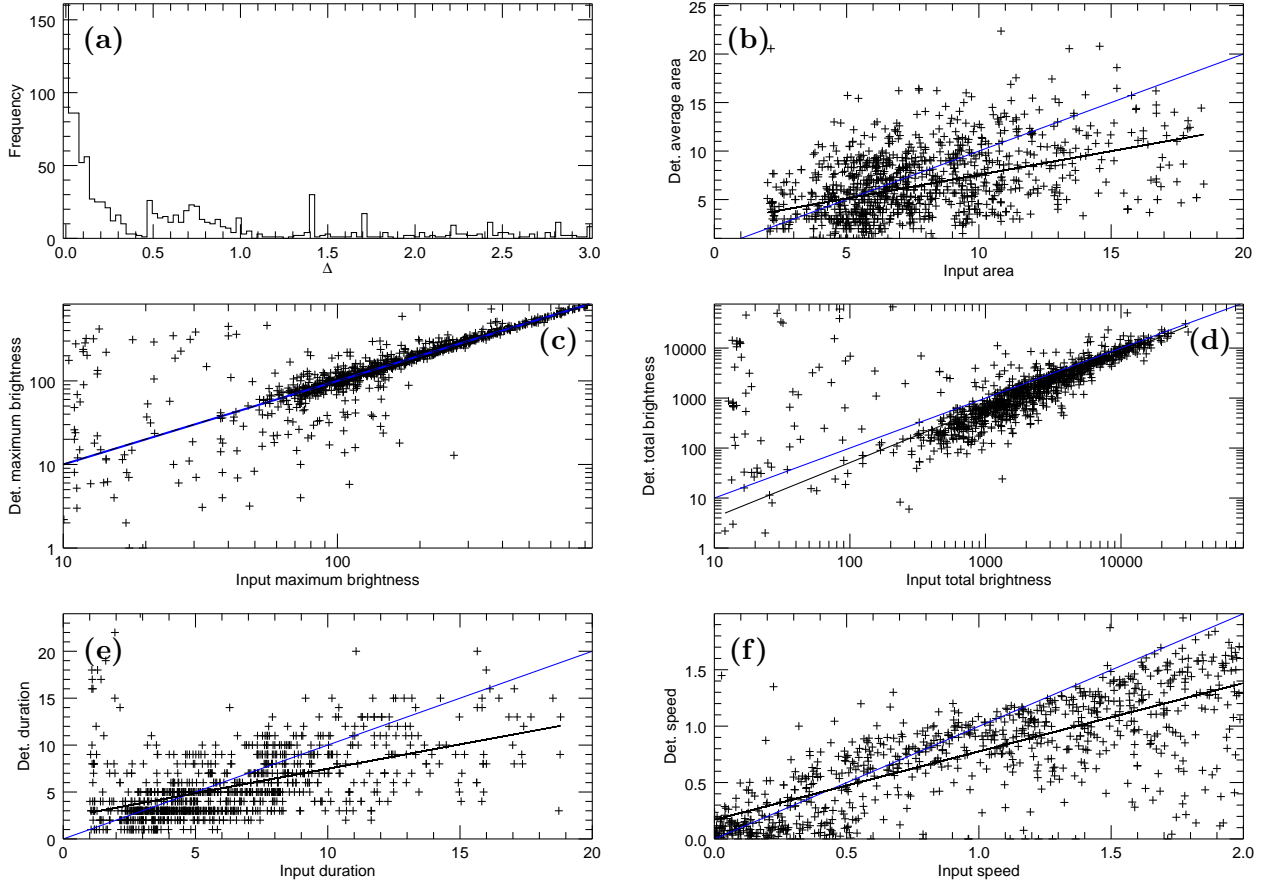}};
    \begin{scope}[
        x={(image.south east)},
        y={(image.north west)}
    ]
        \node [black, font=\bfseries] at (0.13,0.91) {(a)};
        \node [black, font=\bfseries] at (0.605,0.91) {(b)};
        \node [black, font=\bfseries] at (0.465,0.59) {(c)};
        \node [black, font=\bfseries] at (0.945,0.59) {(d)};
        \node [black, font=\bfseries] at (0.13,0.26) {(e)};
        \node [black, font=\bfseries] at (0.605,0.26) {(f)};
    \end{scope}
\end{tikzpicture}
\caption{Plots comparing input and detected properties for the $\approx$1100 paired input-output events (see text). (a) Distribution of $\Delta$ values (or spatio-temporal distance), (b) detected vs. true average area, (c) detected vs. true maximum brightness, (d) detected vs. true total brightness, (e) detected vs. true duration and, (f) detected vs. true speed.
The blue lines indicate one-to-one relationships while the black line is a least-absolute-deviation (LAD) fit to the points. Note that the LAD fit is done in log space for the brightness of (c) and (d), with the others fitted in linear space. The black and blue lines in (c) overlap.}
\label{fig:10k}
\end{figure*}

Figure \ref{fig:10k} shows a one-to-one comparison of input and detected events. The distance (or $\Delta$, as defined previously) between all input events and a detection is calculated, and the input event with the smallest $\Delta$ is paired to that detection. For comparison, we only consider the $\sim11\%$ of input events with $\Delta \leq 3$. The distribution of $\Delta$ for these selected pairs is shown in figure \ref{fig:10k}a. To statistically compare the detected and input properties, we fit the scattered points to a straight line using a least-absolute-deviation (LAD) fitting procedure, and calculate the median relative deviation (MedRD, which will be positive/negative for systematic over/underestimation) and the median absolute relative deviation (MedRAD, which is the spread of deviations from the MedRD). The LAD fitted lines are shown in black in figures \ref{fig:10k}b-f. Note that the fitting is done in log space for \ref{fig:10k}c and d, and in linear space otherwise. The blue lines in \ref{fig:10k}b-f indicate the exact one-to-one relationship.

Figure \ref{fig:10k}b compares the input and detected areas for the paired brightenings. The detected areas tend to underestimate the true areas with a MedRD of $-15$\%, with an increasing underestimation with increasing area shown by the LAD fit gradient of 0.5. The scatter of points is high, with a MedRAD of 30\%.
The true log brightness shown in figure \ref{fig:10k}d shows an overwhelming underestimation of log brightness for the bulk of events with a MedRD value of $-30\%$. The gradient of 1.1 shows that the larger/brighter events give a better estimate of total brightness, with a large scatter for dim/small events and an overall MedRAD of 22\%. 

The durations shown in figure \ref{fig:10k}e are similar to the areas - a tendency to underestimate (MedRD of $-15\%$), with the underestimation increasing with increasing duration (LAD gradient of 0.5), and a large scatter (MedRAD of 25\%).
Figure \ref{fig:10k}f shows the performance of the method in tracking the centroid of events over time through comparing average speeds. The comparison is reasonable given the challenging task. The detected speeds increasingly underestimates the speed with increasing speed, with a gradient of 0.6, a MedRD of $-16\%$ and a MedRAD of 23\%. 

\begin{figure*}[t]
\centering
\includegraphics[trim={0cm 0cm 1cm 0cm},clip,width=\textwidth]{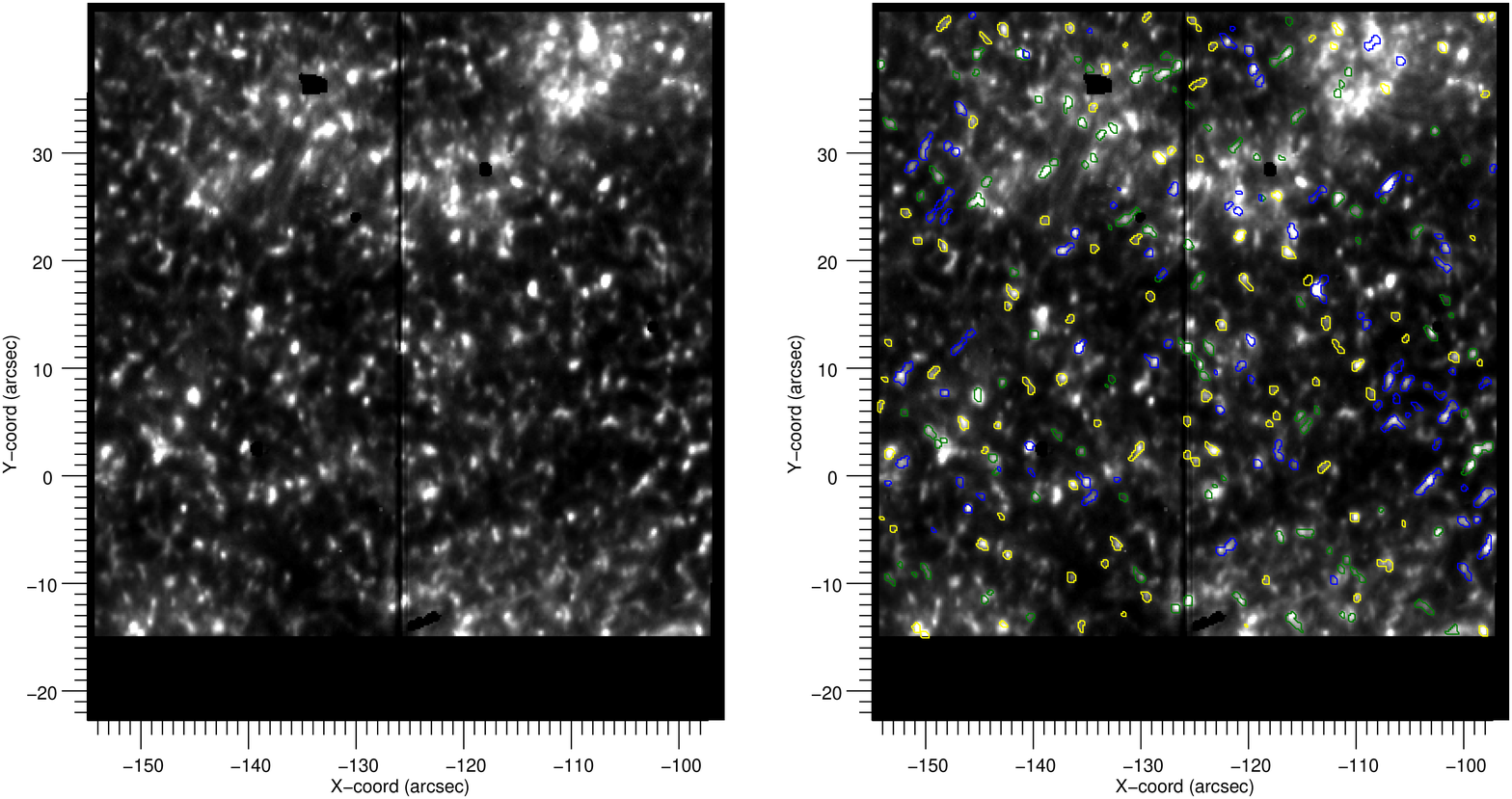}
\caption{IRIS 1400 \AA$\,$ slit-jaw images of (left) the ROI with (right) an over-plot of all detections from the same time frame. The left image has been processed such that isolated pixels or small regions of outlying intensity (such as the central slit) are treated as missing data. 
An animation of this figure is available. The video begins on Oct. 5th 2013 at 08:17 UT. The video ends the same day at 08:36.
}
\label{fig:im_comp}
\end{figure*} 

These comparisons will be central in interpreting the results, and power laws in particular, from real data in future studies, and highlight the difficulty of analysing power laws of small-scale brightenings in solar data. The one-to-one comparisons (only possible of course on synthetic data) show that only the detected maximum brightness can be considered a reliable measure for the brightest and most accurately detected events with a low uncertainty of $\approx$6\%. With real data, one way of confirming whether a detection is true or not is to calculate $\Delta$ across several wavelength channels. If an event appears across several channels and these facets are within an acceptable proximity to each other then the event is deemed to be true and the maximum brightness values recorded for those events are reliable. This will be implemented in a subsequent paper.
The other values are mainly useful for showing trends or relationships for large statistical samples, with a systematic underestimation of area, duration, and total brightness, and a general statistical uncertainty of around 30\%. 
%------------------------------------------------- Observations section --------------------------------------------

\begin{figure*}[t]
\centering
\includegraphics[trim={0 0 0 0},clip,width=0.6\textwidth]{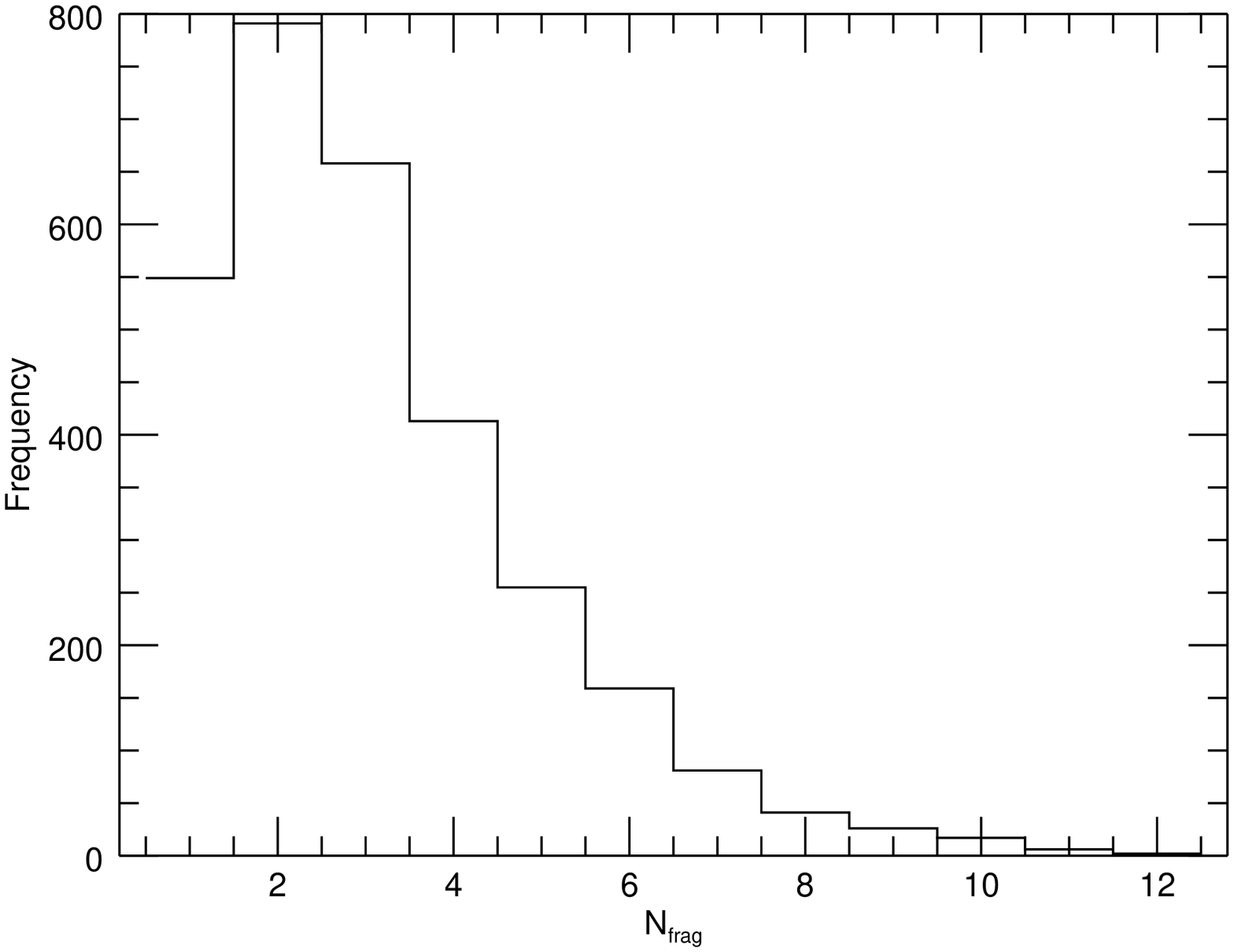}
\caption{Histogram of fragmentation during the lifetime of all detections. 549 events remain un-fragmented with $N_{frag}=1$, 791 detections fragment once with $N_{frag}=2$, 658 events have $N_{frag}=3$, 414 events have $N_{frag}=4$, and the remaining 585 detections have $N_{frag}$ values between 3 and 12.
}
\label{fig:nfrag}
\end{figure*}

\begin{figure*}[t]
\centering
\includegraphics[trim={0cm 0 0.8cm 0.0cm},clip,width=\textwidth]{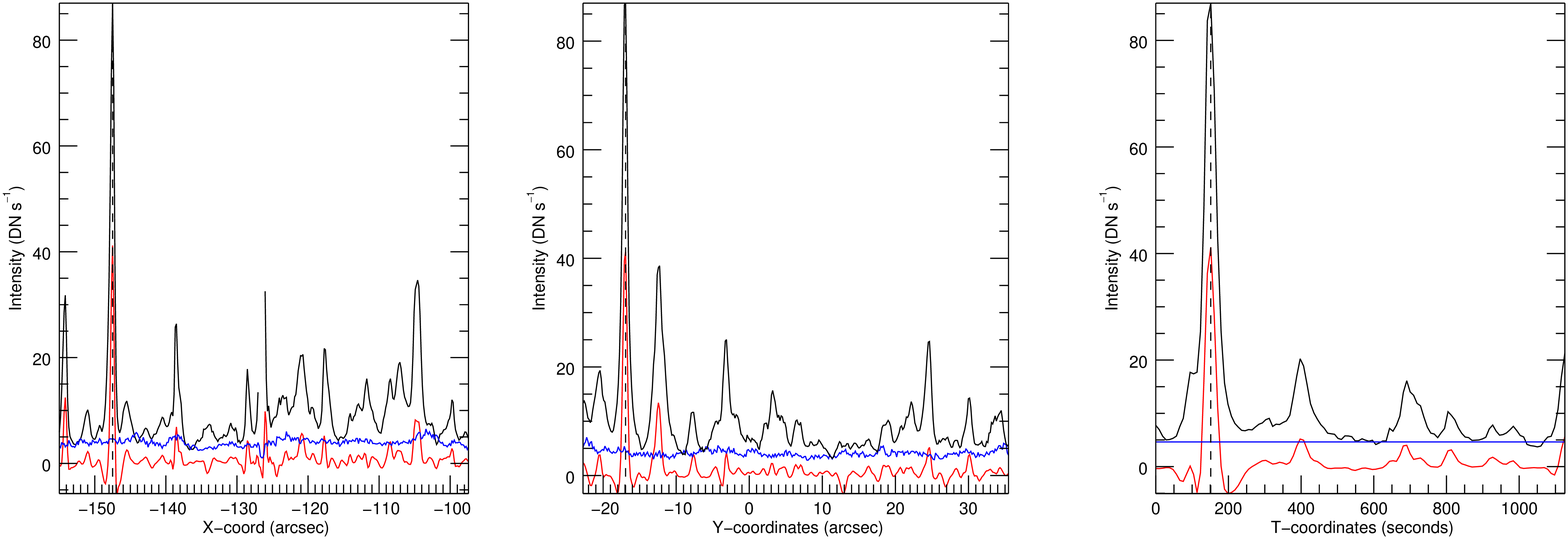}
\caption{An example of the filtering process on slices taken through the IRIS observation datacube's (left) spatial $x$, (center) spatial $y$, and (right) temporal $t$ dimensions, passing through a point coinciding with an event detection, denoted by the vertical dashed line. The black line is the level-2 IRIS signal, the blue denotes the threshold, and red is the filtered data. Intensity values are measured in DN s$^{-1}$.
}
\label{fig:1D_comp_real}
\end{figure*}

\subsection{IRIS data} \label{sec:results real}

We analyse one data set from IRIS \citep{pontieu} centred at approximately X = $-126.05\arcsec$, Y = $5.87\arcsec$. 
The observations begun on Oct. 5th 2013 at 08:17 UT and ends the same day at 08:36. 
Figure \ref{fig:im_comp} (left) shows the region of interest (ROI) as a QS region.
The IRIS data consist of a 1400 \AA\, sit-and-stare series of slit-jaw images, with a consistent 9.46 s temporal cadence, 0.17$\arcsec$ pixel$^{-1}$ spatial scale and a $57.89\arcsec\times58.56\arcsec$ field of view (FOV).

The IRIS data set is processed using standard procedures to account for temporal exposure, and empty margins are removed. 
Dark current subtraction as well as flat field, geometrical and orbital variation corrections have been applied as standard procedure for level 2 IRIS data.
Isolated pixels with spuriously high values are flagged as missing data, and are ignored in the filtering and detection method.
The central slit and its surrounding pixels are also treated as missing data such that extremely `dark' pixels do not affect the filtering process. Additionally, drift due to solar rotation has been treated using a Fourier Local Correlation Tracking method \citep{fw}.

\begin{figure*}[t]
\centering
\includegraphics[trim={0.5cm 0 0.5cm 0.0cm},clip,width=\textwidth]{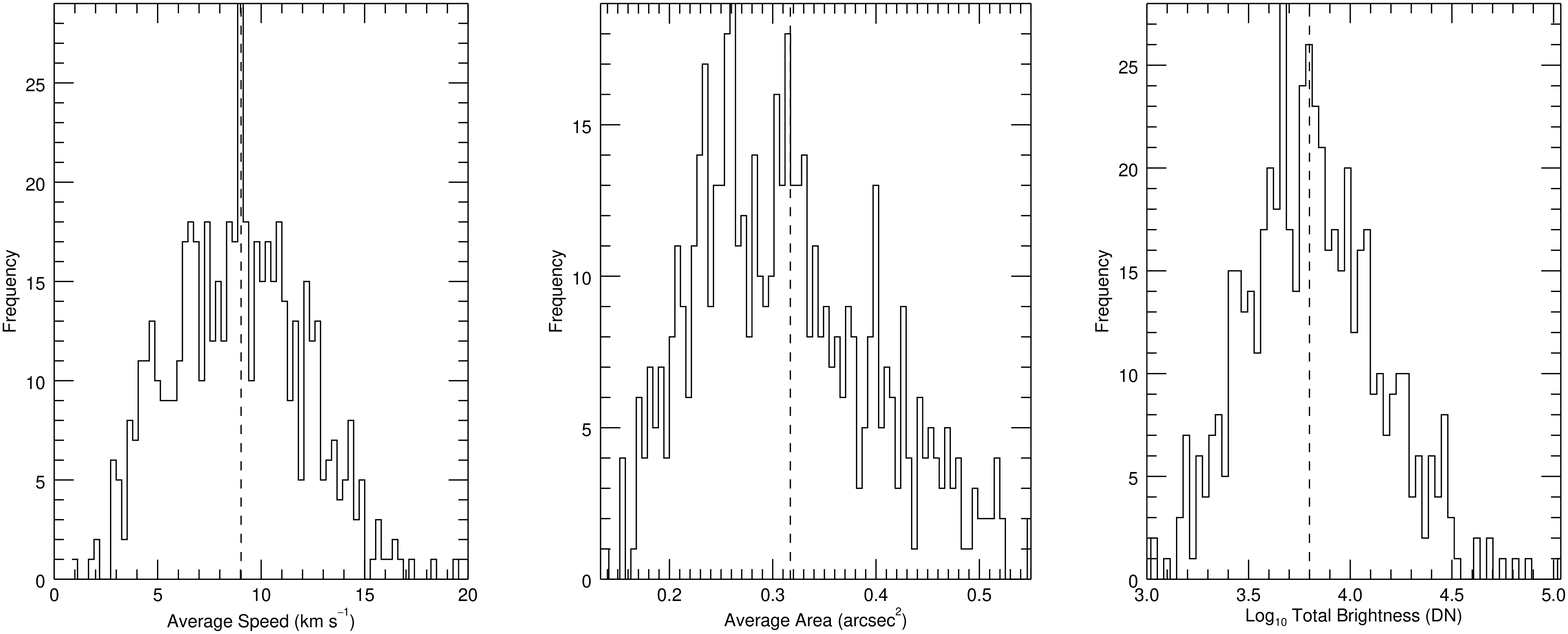}
\caption{Basic properties of the detections from the Oct. 5th 2013 IRIS data. The left, centre and right panels plot histograms of average speed, average area and total brightness, respectively. The dotted vertical lines represent the mean of each property; 9.03 kms$^{-1}$, 0.32 arcsec$^{2}$ and 6300 DNs$^{-1}$ for speed, area and total brightness, respectively.
}
\label{fig:hists}
\end{figure*}

%------------------------------- Results & Discussion section ----------------------------------

The results of the detection method with the IRIS datacube can be seen in figure \ref{fig:im_comp}, whereby the left plot represents the first image of the IRIS datacube. The right plot is the same IRIS image with all brightenings detected in the same time frame as the image on the left. Yellow contours represent brightenings with $N_{frag}\le2$, green represent $N_{frag}=3$ or $N_{frag}=4$, and blue represents $N_{frag}\ge5$. $2997$ small-scale brightening events are detected within the $54.40\arcsec\times55.23\arcsec$ FOV and 18.76-minute observation period, whereby $549$ detections correspond to $N_{frag}=1$ and 791 detections correspond to $N_{frag}=2$, as can be seen in figure \ref{fig:nfrag}. This yields a combined event density of $3.96\times10^{-4}$ events per second per arcsec$^{2}$. The remainder consist of $N_{frag}$ values between 3 and 12. This distribution of fragmented detections is quite different to the test datacube of section \ref{section:10k}, and shows that the brightening detections in IRIS data tend to be more incoherent and fragmented. A more detailed analysis can show whether this is due to a true property of brightenings, or due to some other effect (e.g. high time variability of the background brightness, on which brightening events are superimposed, or overlapping of events in high-activity regions). For the preliminary results presented here, we have chosen to only include brightenings with $N_{frag}=1$.

The choice of $f_{low}$ and $f_{high}$ has substantially reduced both small-scale, rapid noise and large-scale, slow changes, as can be seen in figure \ref{fig:1D_comp_real}. The vertical dashed lines represent the coordinates of a real detection according to its maximum brightness values. It is clear from these plots, similar plots made for other detection events, as well as values of $\Delta$ for synthetic data, that the method is effective at identifying and accurately determining the location of detections.

Figure \ref{fig:hists} presents some preliminary statistical results of these detections's properties: average speed (left), spatial area (center) and total brightness (right). Most detections have a speed between 3 and 15 kms$^{-1}$ (with a mean value of 9.03 kms$^{-1}$), cover an approximate average area between 0.2 and 0.4 arcsec$^{2}$ (with a mean value of 0.32 arcsec$^{2}$) and have a total brightness of $1600-31600$ DN (with a mean value of $6300$ DN). Large and/or bright events are rarer than small and/or dim events, and the distributions are all approximately symmetrical. 
As mentioned in section \ref{section:10k}, the detection process is likely limited by observational resolution which would explain the distribution drop-off for smaller/dimmer/slower events.

Figure \ref{fig:hists} (right)'s profile, above the mean log brightness of $\sim$3.8, suggests that these brightenings follow a power-law distribution. Distributions such as these will be studied in more detail in future work.

\begin{figure*}[t]
\centering
\includegraphics[trim={2cm 0 1.2cm 0.0cm},clip,width=\textwidth]{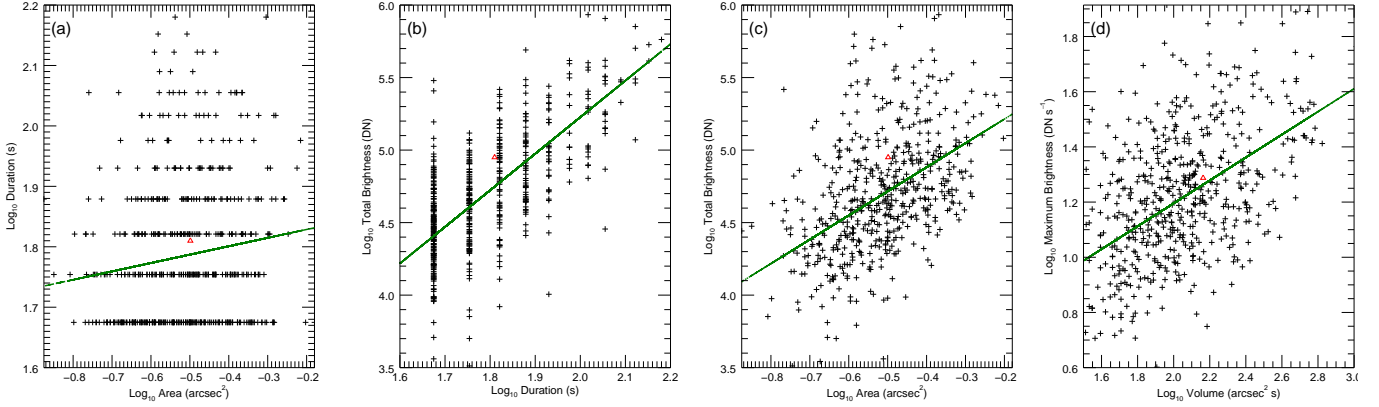}
\caption{Log-log scatter plots showing relationships between (a) area and duration, (b) duration and total brightness, (c) area and total brightness, and (d) volume vs maximum brightness. The green lines represent a robust linear fit to all points. The red triangle shows the mean value of both variables.
}
\label{fig:real_results}
\end{figure*}

\begin{figure*}[t!]
\begin{tikzpicture}
    \node[anchor=south west,inner sep=0] (image) at (0,0) {\includegraphics[trim={2cm 0cm 0cm 0cm},clip,width=\textwidth]{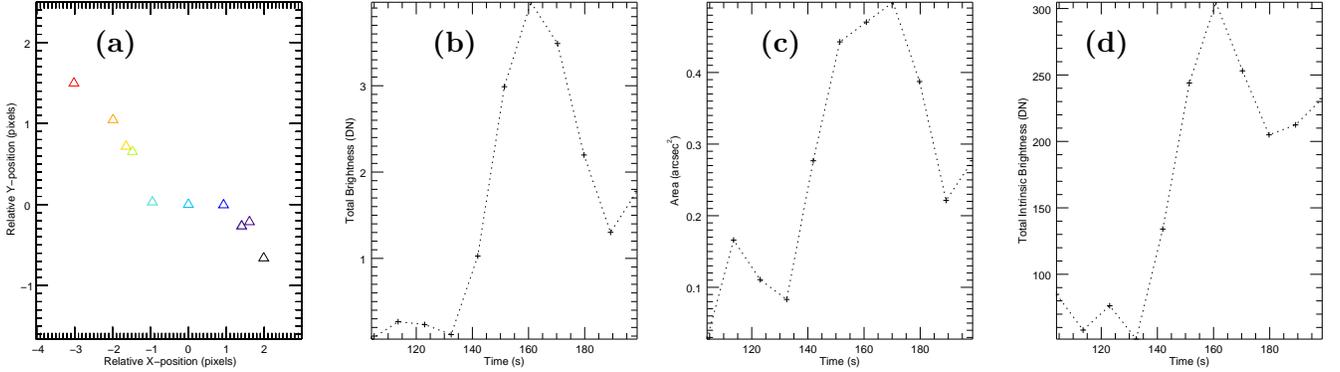}};
    \begin{scope}[
        x={(image.south east)},
        y={(image.north west)}
    ]
        \node [black, font=\bfseries] at (0.09,0.86) {(a)};
        \node [black, font=\bfseries] at (0.34,0.86) {(b)};
        \node [black, font=\bfseries] at (0.58,0.86) {(c)};
        \node [black, font=\bfseries] at (0.82,0.86) {(d)};
    \end{scope}
\end{tikzpicture}
\caption{Moment-by-moment properties of a single brightening event, including (a) the movement of the brightening during its lifetime (from black to red), (b) the change in total brightness over time (measured in DN), (c) the change in area over time (measured in arcsec$^{2}$), and (d) the change of total intrinsic brightness per pixel (measured in DN).
}
\label{fig:step}
\end{figure*}

Figure \ref{fig:real_results} shows log-log scatter plots of area vs duration (a), total brightness vs duration (b), total brightness vs area (c) and volume vs maximum brightness (d). The green line shows the result of a robust linear fit to the logarithmic values, giving gradients of 0.14 (a), 1.60 (b), 1.32 (c) and 0.45 (d). The red triangle represents the mean value of both variables within each plot. The shallow gradient in figure \ref{fig:real_results}a between area and duration suggests that there is no relationship between these properties. However, this result may be limited by the IRIS temporal cadence; the detection process detects the duration of events in integer number of frames, which can be seen by the sudden jumps in values along the y-axis. The plots of figures \ref{fig:real_results}b-d show a clear proportional relationship between duration with total brightness, area with total brightness, and volume with maximum brightness, respectively. The former two relationships are unsurprising considering that a measurement of total brightness is dependent on the number of voxels. The latter relationship suggests that larger events tend to be intrinsically brighter than smaller events.

Figure \ref{fig:step} shows the power of the method in analysing the properties of a brightening over its lifetime. Figure \ref{fig:step}a tracks the motion of the brightening, which has an average speed of $0.06$ pixel s$^{-1}$. 
%$7.66$ kms$^{-1}$
Figure \ref{fig:step}b shows the change in total brightness over time (measured in DN). Figure \ref{fig:step}c shows the change in area over time. Figure \ref{fig:step}d shows the change of maximum brightness over time. We find that this profile is typical of a large number of brightenings which have the property \nfrag$=1$, with a coherent motion and smooth increases/decreases in area and brightness during the event's lifetime. While increases in brightness and area intuitively coincide, figure \ref{fig:step}d suggests that the intrinsic brightness also increases and decreases (independent of the size of the event).

A more detailed analysis of these relationships, applied to this and other data sets, will be made in a future study. We can draw comparisons between these and other studies of small-scale brightening phenomena, such as: H$\alpha$ Ellerman Bombs \citep{ellerman}, which demonstrate typical widths, lengths and durations of $0.63\arcsec$, $0.35\arcsec\,$ and $120s$, respectively \citep{nelson_2}; UV IRIS bombs of approximately $2\arcsec\times2\arcsec$ area \citep{chen}, and; \ion{Ca}{2} chromospheric brightenings lasting at least 1.5 minutes over a $2\arcsec\times2\arcsec$ area \citep{kuckein}. Additionally, figures \ref{fig:hists} (left) and \ref{fig:hists} (centre) demonstrate similar distributions to those of \cite{hou_2016}'s smaller area results and \cite{tian}'s shorter duration results, respectively. Perhaps the results of this paper describe \cite{vissers}'s Flaring Arch Filaments rather than larger and longer-duration phenomena, which may explain the motion of some of these brightenings. We note that most of these previous studies focus on spectroscopic analyses which will lead to differences between their results and those of this study.

%------------------------------------ Conclusions and Future Work section ----------------------------------------
\section{Conclusions and Future work} \label{sec:disc}
We present a general method to detect small-scale brightenings in solar time-series imagery. The method is applied to IRIS data sets in the UV, but can be applied to any time-series image data given appropriate adjustment of a few parameters.
For a simple test dataset of IRIS-like synthetic slit-jaw image time series with randomly-distributed, randomly-moving small-scale Gaussian brightenings, the filtering and detection method yields an event-by-event detection rate of $\sim81\%$ and a voxel-by-voxel detection rate of $\sim49\%$ (with a voxel-by-voxel FP rate of $<10\%$). The detected maximum brightness values provide good accuracy to the true values for the brightest and most accurately detected events. The speeds are also reliable, albeit with a higher scatter. Care must be taken with measurements of total brightness - our tests show a large underestimation (0.3 on average) of total brightness due to the number of fainter voxels not being detected.
The method is also applied to a second datacube with a high density of events possessing a power law distribution of brightness, area and duration. This test demonstrates the difficulty of extracting power law properties from real data due to instrument and method sensitivities to a limited range of brightness, duration and area. However, even though detected properties can have a large scatter from the true values, the shape of the detection histograms largely match the original distributions over a certain range. It also shows that the method gives an estimate of event maximum brightness with low uncertainty. Other properties (area, duration, speed, total brightness) have a large uncertainty, but the method gives approximate proportional relationships over large statistical samples.

2997 brightenings are detected in real IRIS data within a $54.40\arcsec\times55.23\arcsec$ area spanning 19 minutes. 1340 of these detected events either remain unfragmented or fragment at least once over their lifetime. These detections yield an approximate event density of $3.96\times10^{-4}$ arcsec $^{-2}$ s$^{-1}$. Most detections have an average speed between 3 and 15 kms$^{-1}$ (with a mean value of 9.03 kms$^{-1}$), cover an approximate area between 0.2 and 0.4 arcsec$^{2}$ (with a mean value of 0.32 arcsec$^{2}$) and emit a total brightness of $1600-31600$ DN (with a mean value of $6300$ DN).
Linear fitting of brightening properties shows that there are proportional relationships between the brightenings' area with total brightness, duration with total brightness, and volume with maximum brightness. However, there seems to be little proportionality between the brightenings' area with duration: this may be limited by IRIS's temporal cadence. Preliminary frame-by-frame plotting of a brightening event demonstrates coherent motion and a change in total brightness, area and intrinsic brightness over its lifetime - this seems typical of a large number of brightenings. 

Having demonstrated the method's performance, we plan to apply it to a larger sample of quiet-sun IRIS data. This future study will use all the IRIS channels, as well as co-aligned data from other EUV imaging instruments, allowing a statistical study of brightening properties across a range of temperatures/heights. We also plan to test the method on ground-based photospheric data taken in visible wavelengths. The Interactive Data Language (IDL) software described in this paper will be released to the community in the near future through the SolarSoft library.

\acknowledgments

We thank an anonymous referee for comments that greatly improved this work. We acknowledge (1) STFC grant ST/S000518/1 to Aberystwyth University which made this work possible; (2) STFC PhD studentship ST/S505225/1 to Aberystwyth University; and (3) a Coleg Cymraeg Cenedlaethol studentship award to Aberystwyth University.
D.K. has received funding from the S\^{e}r Cymru II scheme, part-funded by the European Regional Development Fund through the Welsh Government and from the Georgian Shota Rustaveli National Science Foundation project FR17 323. 
IRIS is a NASA small explorer mission developed and operated by LMSAL with mission operations executed at the NASA Ames Research center and major contributions to downlink communications funded by ESA and the Norwegian Space Centre. 
The authors wish to acknowledge FBAPS, Aberystwyth University and SuperComputing Wales for the provision of computing facilities and support.

\vspace{5mm}
%\facilities{}
\bibliography{paper2_llh18}
\bibliographystyle{aasjournal}

\end{document}